\documentclass[aps,prl,reprint,groupedaddress,showpacs]{revtex4-1}

\usepackage{amsmath}
\DeclareMathOperator{\Tr}{Tr}
\usepackage{amssymb}
\usepackage{amsthm}

\newtheorem*{theoremNN}{Theorem}

\usepackage{braket}

\usepackage{graphicx}

\usepackage{color}

\begin{document}


\title{Statistical physics of directional, stochastic chains with memory}


\author{J. Ricardo Arias-Gonzalez}
\email[Corresponding author: ricardo.arias@imdea.org]{}
\affiliation{Instituto Madrile\~{n}o de Estudios Avanzados en Nanociencia,
C/Faraday 9, Cantoblanco, 28049 Madrid, Spain}
\affiliation{CNB-CSIC-IMDEA Nanociencia Associated Unit
``Unidad de Nanobiotecnolog\'{i}a"}


\date{\today}

\begin{abstract}
Stochastic chains represent a wide and key variety of phenomena
in many branches of science within the context of Information Theory and
Thermodynamics.
They are typically approached by a
sequence of independent events or by a memoryless Markov process.
Here, we demonstrate that when memory is introduced,
the statistics of the system depends on the mechanism
by which objects or symbols are assembled, even in the slow dynamics limit
wherein friction can be neglected.
To analyse these systems, we introduce a sequence-dependent
partition function, investigate its properties and compare it to
the standard normalization defined by the statistical physics of ensembles.
Then, we study the behaviour of the entropy and the internal energy in this
intrinsic, directional chains finding that they vary around
their thermally-induced equilibrium analogues due to memory effects.
We anticipate that our results are necessary to interpret configurational
order and information transfer in many molecular systems
within Materials Science, Biophysics, Communication and Engineering.
\end{abstract}


\maketitle


Many processes in nature are directional. For example, replication,
transcription and translation in Biology involve molecular machines that
polymerize individual molecular subunits into linear chains in only one
direction out of two~\cite{Arias-Gonzalez2014,Bustamante2011}.
These systems are paradigmatic of the relation between
thermodynamic and information entropy because the chaining of nucleotides or
aminoacids according to a template inherently carries genetic information from
which species propagate~\cite{Berut2012,Landauer1961}.
More in depth, an entropic
configuration of molecular subunits involves a symbol sequence that
conveys information from DNA to proteins.
Specific sequences are recognized by specialized proteins that can not only
correlate present subunit insertion but also proofread and edit errors, a
process that is possible thanks to the existence of memory.
Language writing, copying, reading and editing are also directional,
linear processes
in which symbols are arranged and correlated with previous ones to make
meaningful sentences~\cite{Shannon1948}.
The existence of memory effects and directionality in
Physics and Information Theory, not only regarding natural but also for the
development of artificial systems, is likewise an essential matter with
profound roots~\cite{Bennett1982,Brandao2008,Liu2011}.

Nanoscale chains are subjected to fluctuations and
therefore, a statistical treatment is necessary to understand their physics.
In equilibrium, linear systems are normally
described by a density operator, as defined by the Boltzmann
exponential of the Hamiltonian normalized to the
trace~\cite{Pathria2011,Chandler1987}.
Linear chains that involve an assembly dynamics are directional arrangements
with history, and therefore these systems implicitly comprise spatial and
temporal aspects.
Memoryless stochastic processes,
namely those whose evolution only depend on present
events, are normally treated by Markovian conditional probabilities and general
non-Markovian processes are treated by memory {\it kernel}
functions and dynamical maps~\cite{Wang1945,Presse2013,Breuer2012,Rivas2014},
which involve phenomenological ansatzs. However, to our knowledge, the full
memory in a linear stochastic chain has not been considered, this problem
being especially important to relate non-Markovianity to stochastic processes
that can be characterized by a partition function.

In the following, we show that stochastic chains with memory that are fueled
in one direction in the absence of friction cannot be treated by the standard
statistical ensemble normalization but by a sequence-dependent partition
function. To do this, we use conditional probabilities of full extent to
previous neighbours for the growing chain, i.e. we consider the full memory of
the chain, thus including correlations over present and all the past events.
For the sake of brevity, the terms {\it memory} and {\it history} will
hereafter refer to both present and past events.

The system, either quantum or classic, is built by ordering objects on a
linear chain~\cite{Cover1991,Chandler1987}. We assume that the
interaction of one object, $i$, with the rest, $1,\ldots,i-1,i+1,\ldots, n$,
is restricted
to its previous neighbours, $1, 2, \ldots,i-1$.
Although the interaction with forward members can exist, we suppose that it
does not affect the sequence construction.

A pure state, $\nu$ (or $\ket{\nu}$ in quantum notation), of the system is
specified by a sequence of objects, 
$x_1,\ldots,x_n$, which stem from a multivaluate random variable $\textbf{X}$,
as denoted by:
\begin{equation}\label{eq:state}
\nu = \left\{ x_1, x_2, \ldots, x_i, \ldots, x_{n-1}, x_n \right\}.
\end{equation}
%
%
\noindent
Values $x_i$ represent symbols from an alphabet,
particles from a set of fixed types,
or physical variables (position, momentum, spin, vibrational frequency,
occupation number, etc.).
Their alphabet or domain will be denoted by ${\cal X}$.
Each random variable is in turn a function of several variables of the
microscopic, local environment,
$\Omega$, as for example, pressure, temperature,
ionic conditions or pH.

The most general Hamiltonian is:
\begin{eqnarray}\label{eq:hamiltonian}
H \left( {\bf X} \right) & \equiv & H \left( X_1,\ldots, X_n \right)
\nonumber \\ [+1mm]
& = &
\sum_{i=1}^{n} H\left( X_i; X_{i-1}, \ldots, X_1 \right),
\end{eqnarray}

\noindent
where, if needed, each term of the expansion,
$H\left( X_i; X_{i-1}, \ldots, X_1 \right)$
can in turn be expanded as:
\begin{eqnarray}\label{eq:hamiltoniani}
& & H\left( X_i; X_{i-1}, \ldots, X_1 \right) = H^{0} (X_i)
\nonumber \\ [+1mm]
& + &
H^{(int)} \left( X_i; X_{i-1}, \ldots, X_1 \right) + \frac{1}{n} V.
\end{eqnarray}

\noindent
$H^{0} (X_i)$ may represent, for example, translational and rotational
fluctuations, but generally speaking these energies are due to phenomena
that take place at position $i$ on the chain;
$H^{(int)} \left( X_i; X_{i-1}, \ldots, X_1 \right)$ is the interaction
Hamiltonian and $V/n$ represents an existing external potential similarly
acting on all the objects. Each object is therefore affected by its previous
neighbours and by external phenomena at position $i$, as $X_i (\Omega)$.  

The energy of state $\nu$ is:
\begin{eqnarray}\label{eq:energy}
E_{\nu} \equiv E \left( {\bf x} \right) & = &
E \left( x_1,\ldots, x_n \right)
\nonumber \\ [+1mm]
& = &
\sum_{i=1}^{n} E\left( x_i; x_{i-1}, \ldots, x_1 \right),
\end{eqnarray}

\noindent
where the {\it partial energy} $E(x_i; x_{i-1}, \ldots, x_1)$
is the energy of object $x_i$ provided that the previous objects
(which constitute a {\it partial sequence})
are $(x_1, \ldots x_{i-1})$. Energies $E_{\nu}$
are formally obtained in the quantum case as eigenvalues of $H$,
$E_{\nu} = \Braket{\nu | H \left( {\bf X} \right) | \nu}$,
where $\ket{\nu} \in \mathcal{H}$,
the corresponding eigenvectors in a Hilbert space, are stationary
solutions of $H\ket{\nu} = E_{\nu}\ket{\nu}$,
with $\Braket{\nu | \nu'} = \delta_{\nu \nu'}$.
Each vector $\ket{\nu}$ is constructed on the basis of the distinguishability
and symmetry of the quantum objects that comprise the chain.

The probability of a state is:
\begin{eqnarray}\label{eq:probability}
p_{\nu} & \equiv & \text{Pr} \{X_1 = x_1, \ldots,  X_n = x_n \} =
p \left( {\bf x} \right) = p(x_1, \ldots, x_n)
\nonumber \\ [+1mm]
& = &
p(x_1)p(x_2|x_1) \cdots p(x_n|x_{n-1}, \ldots, x_1),
\end{eqnarray}

\noindent
where the last part of the equation is the general expansion of the joint
probability as a product of conditional probabilities~\cite{Fisz1980}.
It can be factorized since the probability at each step, $i$, depends only
on the previous neighbours.

The Gibbs entropy of the system is:
\begin{eqnarray}\label{eq:entropy}
& S &(X_1, \ldots, X_n) =
-k \left\langle \ln p \right\rangle  = - k \sum_{\nu=1}^N p_{\nu} \ln p_{\nu}
\nonumber \\ [+1mm]
& = &
- k \sum_{x_1, \ldots, x_n} p(x_1, \ldots, x_n) \ln p(x_1, \ldots, x_n),
\end{eqnarray}

\noindent
where $k$ is the Boltzmann constant, ``$\ln$" the natural logarithm and
$N$, the number of microstates.
The mean (internal) energy of the system is:
\begin{equation}\label{eq:MeanEnergy}
\left\langle E \right\rangle = \sum_{\nu=1}^N p_{\nu} E_{\nu} =
\sum_{x_1, \ldots, x_n} p \left(x_1, \ldots, x_n \right)
E \left( x_1, \ldots, x_n \right),
\end{equation}

\noindent
where $N$ is the number of microstates.

In equilibrium with balanced flows of matter or energy, probabilities can
be normalized to a partition function in the form:
\begin{eqnarray}\label{eq:Z}
Z (\beta, n) & \equiv & \sum_{\nu=1}^{N} \exp{\left[ -\beta E_{\nu} \right]}
\nonumber \\ [+1mm]
& = &
\sum_{x_1, \ldots, x_n} \exp{\left[ -\beta E\left( x_1, \ldots, x_n \right)
\right]}
\nonumber \\ [+1mm]
& = &
\sum_{x_1, \ldots, x_n} \exp{\left[ -\beta \sum_{i=1}^{n}
E\left( x_i; x_{i-1}, \ldots, x_1 \right) \right]},
\end{eqnarray}

\noindent
where $\beta = 1/ k T$ and $T$ is the absolute
temperature. The probability of a configuration is thus:
\begin{equation}\label{eq:ProbIsing}
p_{\nu} = \frac{\exp{\left( -\beta E_{\nu} \right)}}{Z}.
\end{equation}

\noindent
This equation represents the well-known equilibrium probability
for a system that has thermalized into a particular configuration.
No driving forces other than those that guide the spontaneous process are
present.
Then, neither a temporal directionality is imposed
(i.e. object at position $i$
could have been incorporated or taken its final value, $x_i$, before or after
that at position $i+1$, $x_{i+1}$) nor a one-by-one incorporation is assumed.
What is more, this formalism involves that objects can fluctuate
non-sequentially among their different values before a final configuration is
reached. This equilibrium, which involves interacting objects,
is susceptible to be solved paralleling the Ising model.

Many processes are, however, directional, like for example DNA replication,
in which a polymerase protein copies a template DNA strand from the 5' to
the 3' end~\cite{Andrieux2008b,Arias-Gonzalez2012}, or English writing,
which takes place from left to right.
Directionality can be imposed by physicochemical constraints or rules,
irrespectively of whether the process is quasistatic
(i.e. with such a slow dynamics that it approaches equilibrium
conditions at each step) or non-equilibrium.
For such directional processes, the probability has to be calculated at each
step as:
\begin{equation}\label{eq:MarkovProb}
p \left( x_i|x_{i-1}, \ldots, x_1 \right) =
\frac{e^{-\beta E\left( x_i; x_{i-1}, \ldots, x_1 \right) }}
{{\sum_{x'_i}} e^{-\beta E\left( x'_i; x_{i-1}, \ldots, x_1 \right)}}.
\end{equation}

\noindent
This mechanism constraints the sequence in which the different configurations
are accessible at every step and therefore, even though each configuration
may need an infinite time to be constructed, this mechanism does not involve
a global thermalization of the system. The states so accessed comprise
all the possible ones, as in Eq.~(\ref{eq:Z}).
These states, or the individual steps to reach them, should not
be described in general as metastable since,
once a configuration has been achieved,
the system may not abandon it spontaneously
if the state is sufficiently stable. However, the total
probability of a configuration is strictly different than that given by
Eq.~(\ref{eq:ProbIsing}) and can be expressed as:
\begin{equation}\label{eq:ProbMarkov}
p^{(D)}_{\nu} = \frac{\exp{\left( -\beta E_{\nu} \right)}}{Z_{\nu}},
\end{equation}

\noindent
where superindex ``D" denotes a directional mechanism and $Z_{\nu}$ is given
by:
\begin{equation}\label{eq:Znu}
Z_{\nu}  (\beta, n) \equiv
\sum_{x'_1, \ldots, x'_n} \exp{\left[ -\beta \sum_{i=1}^{n}
E\left( x'_i; x_{i-1}, \ldots, x_1 \right) \right]}.
\end{equation}

\noindent
We can name $Z_{\nu}$ {\it sequence-dependent partition function} since it
depends on previous events (not primed variables in Eq.~(\ref{eq:Znu})).
The fact that $\sum_{\nu} p_{\nu} ^{(D)} = 1$ follows from the application of
the fact that $\sum_{x_i} p \left( x_i|x_{i-1}, \ldots, x_1 \right) =1$,
see Eq.~(\ref{eq:MarkovProb}), into Eq.~(\ref{eq:probability}).

The partition function is not therefore unique for directional processes
and this represents the main message of this article, the consequences
and physical meaning of which will be analysed below.

If we neglect interactions with previous neighbours,
Eqs.~(\ref{eq:Z}) and~(\ref{eq:Znu}) converge to the same
value and probabilities Eqs.~(\ref{eq:ProbIsing})
and~(\ref{eq:ProbMarkov})
collapse into the same expression. In other words, in the absence of
previous-neighbour interactions, directionality and global thermalization
into equilibrium are equivalent. This was shown for DNA replication
elsewhere~\cite{Arias-Gonzalez2012} and will be rigorously demonstrated
later on this article.

It is easy to demonstrate the following properties for the sequence-dependent
partition function:
\begin{eqnarray}\label{eq:meanZ}
      \left\langle \frac{1}{Z_{\nu}} \right\rangle = \frac{1}{Z},
      \,\,\,\,\,\,\,
      \left\langle Z_{\nu} \right\rangle_D = Z,
\\ [+1mm]
\label{eq:ineq}
      \sum_{\nu} Z_{\nu}  \geq Z,
      \,\,\,\,\,\,\,
      \exp \left( -\beta E_{\nu} \right) \leq Z_{\nu},
\end{eqnarray}

\noindent
where subindex ``D" in Eq.~(\ref{eq:meanZ}) indicates that the mean
has been taken by using the directional probability given by
Eq.~(\ref{eq:ProbMarkov}) whereas the absence of this subindex
indicates the use of the probability given by Eq.~(\ref{eq:ProbIsing}).
The proof of, for example, Eq.~(\ref{eq:meanZ}) is:
$1=\sum_{\nu} p_{\nu}=(1/Z) \sum_{\nu}Z_{\nu}p^{(D)}_{\nu} =
\left\langle Z_{\nu} \right\rangle_D/Z$.
From Eq.~(\ref{eq:meanZ}), it is easy to
understand that $Z_{\nu}/Z$ can be greater, equal or smaller than 1;
likewise, that $\left\langle p_{\nu}^{(D)}/p_{\nu} \right\rangle =1$ and that
$\left\langle p_{\nu}/p_{\nu}^{(D)} \right\rangle_D =1$.   
 
To continue, we need to introduce the {\it two-sequence Hamiltonian},
$\hat{H}$, as:
\begin{equation}\label{eq:2seqH}
\hat{H} \left( {\bf X', X} \right) \equiv
\sum_{i=1}^{n} H\left( X'_i; X_{i-1}, \ldots, X_1 \right),
\end{equation}

\noindent
whose mathematical expression fulfills
$\hat{H} \left( {\bf X, X} \right) \sim H \left( {\bf X} \right)$.
This Hamiltonian can be introduced in Quantum Statistics
as a sesquilinear form,
$\hat{H}: \mathcal{H} \times \mathcal{H} \rightarrow \mathbb C$, namely,
$(\bra{\mu'} + \bra{\nu'}) \hat{H} (\ket{\nu} + \ket{\mu}) =
                   \bra{\mu'} \hat{H} \ket{\nu} +
                   \bra{\mu'} \hat{H} \ket{\mu} +
                   \bra{\nu'} \hat{H} \ket{\nu} +
                   \bra{\nu'} \hat{H} \ket{\mu}            
$,
and
$\bra{(a) \nu'}\hat{H} \ket{(b)\nu} = \overline{a}b\bra{\nu'}\hat{H} \ket{\nu}$,
for all $\ket{\mu}, \ket{\mu'}, \ket{\nu}, \ket{\nu'} \in \mathcal{H}$
and all $a$ and $b$ $\in \mathbb C$,
being $\overline{a}$ the complex conjugate of $a$, in contrast to $H$, which is
a linear operator $H : \mathcal{H} \rightarrow \mathcal{H}$.
The complex bilinear form $\hat{H}$ does not represent an
observable, the energy operator was actually given in
Eq.~(\ref{eq:hamiltonian}). The matrix elements of $\hat{H}$ are: 
\begin{equation}\label{eq:2seqHelements}
\Braket{\nu' | \hat{H} \left( {\bf X', X} \right) | \nu} = E_{\nu' \nu},
\end{equation}

\noindent
where $E_{\nu' \nu}$ is the {\it two-sequence energy}:
\begin{equation}\label{eq:2seqEnergy}
E_{\nu' \nu} \equiv \sum_{i=1}^{n} E\left( x'_i; x_{i-1}, \ldots, x_1 \right),
\end{equation}

\noindent
and $\nu' = \left\{ x'_1, x'_2, \ldots, x'_i, \ldots, x'_{n-1}, x'_n \right\}$.
Then, the sequence-dependent partition function,
Eq.~(\ref{eq:Znu}), can be spelled as:
\begin{eqnarray}\label{eq:Znu2}
Z_{\nu} & \equiv &
\sum_{\nu'=1}^{N} \exp{\left( -\beta  E_{\nu' \nu} \right)}.
\end{eqnarray}

To fully understand the physical meaning and relationship between the two
partition functions, we represent their terms in matrix form:
\begin{eqnarray}\label{eq:Zmatrix}
\textbf{Z}  & = & \left( Z_{\nu' \nu}\right) =
\left( e^{ -\beta E_{\nu' \nu} }\right)
\nonumber \\ [+2mm]
& =  &
\left(\begin{array}{cccc}
  e^{-\beta  E_{1} } &  e^{ -\beta  E_{12}} 
& \ldots & e^{ -\beta  E_{1N}}\\
e^{-\beta  E_{21}} &  e^{-\beta  E_{2}} 
& \ldots & e^{-\beta  E_{2N}}\\
\vdots & \vdots & \ddots & \vdots\\
e^{-\beta  E_{N1}} &  e^{-\beta  E_{N2}} 
& \ldots & e^{-\beta  E_{N}}
\end{array}\right),
\end{eqnarray}

\noindent
where we have used that $E_{\nu \nu}= E_{\nu}$. This matrix is easily
defined in the Quantum Statistics formalism as:
\begin{eqnarray}\label{eq:ZnuQuantum}
Z_{\nu' \nu} = \Braket{\nu' | e^{-\beta \hat{H}
\left( {\bf X', X} \right)} | \nu},
\end{eqnarray}

\noindent
with which the standard partition function conserves its quantum definition,
\begin{eqnarray}\label{eq:ZQuantum1}
Z = \Tr \textbf{Z} & = &
\Tr \left\{\exp \left[-\beta \hat{H} \left( {\bf X', X}\right)\right]
\right\}
\nonumber \\ [+2mm]
& = &
\label{eq:ZQuantum2}
\Tr \left\{\exp \left[-\beta H \left( {\bf X}\right)\right] \right\}.
\end{eqnarray}

The formalism introduced in
Eqs.~(\ref{eq:MarkovProb}-\ref{eq:ZQuantum2}) allows the
sequence-dependent partition function, Eq.~(\ref{eq:Znu}), to be
integrated into the Statistical Physics framework.
In short, a {\it directional} mechanism in a quasistatic linear chain
construction is introduced by defining probabilities
(Eq.~(\ref{eq:ProbMarkov}))
whose normalization (Eq.~(\ref{eq:Znu})) depends on their history.
Such normalization is obtained by summing up over the column elements on the
matrix representation of the two-sequence Hamiltonian
(Eq.~(\ref{eq:2seqH})). It is noticeable that the density operator, $\rho$
for quantum linear, stochastic chains with memory does not conserve the trace
$\Tr \rho=1$,
but the sum over the elements of each individual column in its matrix
representation,
which is a consequence of the saturation of the conditional probabilities,
$\sum_{x_i} p \left( x_i|x_{i-1}, \ldots, x_1 \right) =1$, that we saw above.

With the aim of providing meaning to the above treatment,
we analyse next the internal energy and the thermodynamic entropy of
such a sequential chain system.
Following Eq.~(\ref{eq:MeanEnergy}), the
internal energy can be expressed as:
\begin{equation}\label{eq:MeanEnergy2}
\left\langle E \right\rangle =
\left\langle \frac{Z_{\nu}}{Z} E_{\nu} \right\rangle_D,
\,\,\,\,\,\,\,
\left\langle E \right\rangle_D =
\left\langle \frac{Z}{Z_{\nu}} E_{\nu}\right\rangle.
\end{equation}

\noindent
Similarly, following Eq.~(\ref{eq:entropy}), the entropy reads:
\begin{eqnarray}\label{eq:entropy2}
S & = &
    - k \left\langle \frac{Z_{\nu}}{Z} \ln \frac{Z_{\nu}}{Z} \right\rangle_D
    - k \left\langle \frac{Z_{\nu}}{Z} \ln p^{(D)}_{\nu} \right\rangle_D,
\nonumber \\ [+2mm]
S^{(D)} & = &
    - k \left\langle \frac{Z}{Z_{\nu}} \ln \frac{Z}{Z_{\nu}} \right\rangle
    - k \left\langle \frac{Z}{Z_{\nu}} \ln p_{\nu} \right\rangle,
\end{eqnarray}

\noindent
where $S^{(D)}$ is the Gibbs entropy of the directional chain,
$S^{(D)} (X_1, \ldots, X_n) = 
-k \left\langle \ln p^{(D)} \right\rangle_D =
- k \sum_{\nu=1}^N p^{(D)}_{\nu} \ln p^{(D)}_{\nu}$.

To gain physical insight, we study next the case in which the chain
construction is sufficiently smoothly-dependent on its history.
First, we formulate and demonstrate the following theorem:

\begin{theoremNN}[Independence limit]\label{th:SeqEnergy}
Let $\nu$ be a stochastic, linear chain with memory
(Eq.~(\ref{eq:state})) sequentially constructed $i:1 \rightarrow n$.
Let $Z$ and $Z_{\nu}$ be the normal (equilibrium) and the sequence-dependent
partition functions (Eqs.~(\ref{eq:Z}) and~(\ref{eq:Znu},\ref{eq:Znu2}),
respectively), and let $E_i= E\left( x_i; x_{i-1}, \ldots, x_1 \right)$ and
$E'_i=E\left( x_i; x'_{i-1}, \ldots, x'_1 \right)$ be the energies of object
$x_i$ relative to two different partial sequences (Eq.~(\ref{eq:energy})).
If the normalized energy difference $|E_i-E'_i| / kT \to 0$, $\forall i$, then
$Z/Z_{\nu} \to 1$.
\end{theoremNN}

This theorem provides the adequate link between directionally-driven
processes and equilibrium thermodynamics in the
absence of friction.
It states that when memory effects are weak enough, the construction of
a stochastic chain by a defined mechanism approaches
an equilibrium thermalization.
{\it Weak enough} here means that the the energetic cost for incorporating any
new object in the growing chain is affected by a sufficiently low quantity
with respect to the thermal energy level ($kT$) due to the sequence of previous
incorporations.

This theorem also applies when sequence-dependent energies, $E_i$ and $E'_i$,
are much lower than the thermal level, $kT$, at every step $i$. Certainly,
$E_i, E'_i \ll kT \Rightarrow \left| E_i-E'_i \right| \ll kT$.
This is the high-temperature limit
in which the thermal energy is so large compared to those associated to
neighbouring interactions that the system dynamics is dominated by random
fluctuations. This is the case of the ideal gas. In particular, following
Eq.~(\ref{eq:hamiltoniani}), the partial energy of a configuration can be
written as:
\begin{eqnarray}\label{eq:energyi}
& E & \left( x_i; x_{i-1}, \ldots, x_1 \right)
\nonumber \\ [+1mm]
& = &
E^{0} (x_i) + E^{(int)} \left( x_i; x_{i-1}, \ldots, x_1 \right) +
         \frac{1}{n} V
\nonumber \\ [+1mm]
& = &
E_i^{0} + E_i + \frac{1}{n} V,
\end{eqnarray}

\noindent
where we have identified $E_i^{0} = E^{0} (x_i)$ and
$E_i = E^{(int)} \left( x_i; x_{i-1}, \ldots, x_1 \right)$.
In these conditions,
$E \left( x_i; x_{i-1}, \ldots, x_1 \right) -
E \left( x_i; x'_{i-1}, \ldots, x'_1 \right) = E_i-E'_i \ll kT$,
and the theorem applies.

\begin{proof}
When memory effects are sufficiently mild, we can express the partial
energies as:
\begin{eqnarray}\label{eq:EdiffLow}
E\left( x'_i; x'_{i-1}, \ldots, x'_1 \right) & = &
E\left( x'_i; x_{i-1}, \ldots, x_1 \right) \nonumber
\\
& + & \epsilon \left(x_{i-1}, \ldots, x_1 \right),
\end{eqnarray}

\noindent
where $\epsilon_i \equiv \epsilon \left(x_{i-1}, \ldots, x_1 \right)$ is
the energy difference between energies associated to different, partial
sequences $(x'_{i-1}, \ldots, x'_1)$ and $(x_{i-1}, \ldots, x_1)$.
For all $i$, it fulfills:
\begin{eqnarray}\label{eq:EpsCond1}
& \epsilon & \left(x_{i-1}, \ldots, x_1 \right) = 0, \nonumber
\\
& & \quad \quad \text{if} \,\,\, 
\left(x_{i-1}, \ldots, x_1 \right) = \left(x'_{i-1}, \ldots, x'_1 \right),
\,\, \text{and} \,\,\,
\\
\label{eq:EpsCond2}
& & \left| \epsilon(x_{i-1}, \ldots, x_1) \right| \ll kT,
\quad \text{otherwise}.
\end{eqnarray}

Then,
\begin{eqnarray}\label{eq:Develop1}
E_{\nu'} & = & \sum_{i=1}^n E(x'_i; x'_{i-1}, \ldots, x'_1)
\nonumber \\ [+1mm]
& = &
\sum_{i=1}^n E(x'_i; x_{i-1}, \ldots, x_1) +
\sum_{i=1}^n \epsilon (x_{i-1}, \ldots, x_1)
\nonumber \\ [+1mm]
& = &
E_{\nu' \nu} + \epsilon_{\nu},
\end{eqnarray}

\noindent
where we have used the definition of the two-sequence energy,
Eq.~(\ref{eq:2seqEnergy}), and that
$\epsilon_{\nu} \equiv \sum_{i=1}^n \epsilon_i$
is a sequence-dependent small energy. Consequently,
\begin{eqnarray}\label{eq:Develop2}
Z-Z_{\nu} & = & \sum_{\nu'=1}^N \exp \left( -\beta E_{\nu'} \right) -
                \sum_{\nu'=1}^N \exp \left( -\beta E_{\nu' \nu} \right)
\nonumber \\ [+1mm]
& = &
                \sum_{\nu'=1}^N \exp \left( -\beta E_{\nu' \nu} \right)
                         \left[ \exp \left (-\beta \epsilon_{\nu} \right)
                               - 1 \right].
\end{eqnarray}

We expand $\exp (-\beta \epsilon_{\nu})$ in the limit
$\beta \epsilon_i \rightarrow 0, \forall i$:
\begin{eqnarray}\label{eq:Develop3}
\exp \left (-\beta \epsilon_{\nu} \right) & = &
\exp \left (-\beta \sum_{i=1}^n \epsilon_i \right) =
\prod_{i=1}^n \exp \left( -\beta \epsilon_i \right)
\nonumber \\ [+1mm]
& \approx & 1 - \beta \sum_{i=1} \epsilon_i = 1 - \beta \epsilon_{\nu}.
\end{eqnarray}

\noindent
From Eq.~(\ref{eq:Develop2}), it follows that
\begin{equation}\label{eq:Develop4}
Z-Z_{\nu} \approx Z_{\nu} \left(- \beta \epsilon_{\nu} \right) =
\varepsilon_{\nu} Z_{\nu},
\end{equation}

\noindent
where $-\beta \epsilon_{\nu} \equiv \varepsilon_{\nu}$ is a small number.
Then, $Z/Z_{\nu} \approx 1 + \varepsilon_{\nu}$
with $\varepsilon_{\nu} \sim 0$.
\end{proof}

As previously stated and further reflected in this demonstration,
the limit in the theorem is reached when interactions with previous neighbours
are neglected; in such a case, $X_i$ can be approximated by independent
random variables.

Let's expand now the internal energy and the entropy,
Eqs.~(\ref{eq:MeanEnergy2}) and~(\ref{eq:entropy2}),
respectively, in the limit
$Z/Z_{\nu} \sim 1$ by setting $Z/Z_{\nu} = 1 + \varepsilon_{\nu}$, being
$\varepsilon_{\nu}$ a real, small number, as defined in the above demonstration.
Under such conditions, the internal energy and entropy,
Eqs.~(\ref{eq:MeanEnergy2}) and~Eq.~(\ref{eq:entropy2}),
respectively, read:
\begin{eqnarray}\label{eq:ApproxE}
\left\langle E \right\rangle_D  & = & \left\langle E \right\rangle +
\left\langle \varepsilon_{\nu} E_{\nu} \right\rangle ,
\\ [+1mm]
\label{eq:ApproxS}
S^{(D)} & = & S
    - k \left\langle \varepsilon_{\nu} \ln p_{\nu} \right\rangle;
\end{eqnarray}

\noindent
the last equation is valid for $|\varepsilon_{\nu}| < 1$,
which is the radius of convergence of the MacLaurin series for
$\ln(1+\varepsilon_{\nu})$.
These expressions state that both the internal energy and the entropy fluctuate
around the values given by the thermal equilibrium process, which comprise
all possible mechanisms in the limit of no friction, when memory
effects are low.

The stochastic chain described as a thermal equilibrium process,
which includes a Boltzmann, scalar partition function, does not involve a
mechanism in the chain construction, and therefore it comprises
all the possible pathways. Directional chain construction, in contrast,
independently of whether it can be approached by an equilibrium stepping
dynamics or not, involves a mechanism that drives the building of the chain
through a certain microscopic pathway.
In both cases, all the possible configurations
are accessible, but in the latter, directionality dictates how the chain must
be constructed.

We have shown that the physics of a stochastic linear chain construction
depends on the existence of a directionality in the presence of memory.
We have demonstrated that configurational probabilities
for thermal equilibrium collapse are strictly different than those for the
directional construction. Then, we have established their thermodynamic
relations, thus interpreting their physical meaning in terms of the internal
energy and entropy.
Directionality correlates with a time arrow, in contrast to
thermal equilibrium, which is timeless. Besides, the fact that the
partition function depends on the history is a consequence of the fact that
a thermodynamic limit cannot be defined for driven stochastic processes.
Therefore, state variables such as the entropy, the mean energy or the free
energy cannot be defined in the thermodynamic limit.
Linear, stochastic chains with memory can be found in diverse, nanoscale,
scenarios. We believe that our results are not only
important to Molecular Biology but also to other more general small systems,
as well as to Computer Science and Communication theory,
because the generation, edition and transfer of information
rely on the use of linear, stochastic chains of
interacting objects, either physical or symbolic. For example,
the composition of human language or the generation of genomes throughout
evolution could be treated under the general scheme described here.

\noindent
\textbf{Acknowledgments}\\
D. G. Aleja is thanked for fruitful discussion.
Work supported by MINECO (grant MAT2013-49455-EXP).

\end{document}